\documentclass{aa}  
\usepackage{color, colortbl}
\usepackage{hyperref}
\usepackage[scale=1]{miama}
\hypersetup{
    colorlinks=true,
    linkcolor=blue,
    citecolor=blue,
    filecolor=black,      
    urlcolor=black,
    }
\usepackage{siunitx}
\usepackage[dvipsnames]{xcolor}
\usepackage{amsmath}
\usepackage{natbib}

\usepackage{graphicx}
\usepackage{txfonts}
\begin{document}

   \title{A combined LOFAR and XMM-Newton analysis of the disturbed cluster PSZ2G113.91-37.01}

   \author{M. G. Campitiello\inst{1,2}\fnmsep\thanks{\email{giulia.campitiello@inaf.it}}, A. Bonafede\inst{1,3}, A. Botteon\inst{3}, L. Lovisari\inst{1,4}, S. Ettori\inst{1,5}, G. Brunetti\inst{3}, F. Gastaldello\inst{6}, M. Rossetti\inst{6}, R. Cassano\inst{3}, A. Ignesti\inst{7}, R.J. van Weeren\inst{8}, M. Br\"{u}ggen\inst{9}, M. Hoeft\inst{9}}

    \institute{INAF, Osservatorio di Astrofisica e Scienza dello Spazio, via Piero Gobetti 93/3, 40129 Bologna, Italy
    \and
    Dipartimento di Fisica e Astronomia, Università di Bologna, Via Gobetti 92/3, 40121, Bologna, Italy
    \and
    INAF – Istituto di Radio Astronomia, Via Gobetti 101, 40129 Bologna, Italy
    \and
    Center for Astrophysics $|$ Harvard $\&$ Smithsonian, 60 Garden Street, Cambridge, MA 02138, USA  
    \and
    INFN, Sezione di Bologna, viale Berti Pichat 6/2, 40127 Bologna, Italy
    \and
     INAF – Istituto di Astrofisica Spaziale e Fisica Cosmica di Milano, Via A. Corti 12, 20133 Milano, Italy
     \and
     INAF-Padova Astronomical Observatory, Vicolo dell’Osservatorio 5, I-35122 Padova, Italy
     \and
     Leiden Observatory, Leiden University, PO Box 9513, 2300 RA Leiden, The Netherlands
    \and
     University of Hamburg, Gojenbergsweg 112, 21029 Hamburg, Germany}
   \date{Accepted 27 September 2023}

 
  \abstract
  {In this work, we investigated the interplay between the X-ray and radio emission of the cluster PSZ2G113.91-37.01 ($z = 0.371$) using the high-quality XMM-Newton observations of the Cluster HEritage project with XMM-Newton - Mass Assembly and Thermodynamics at the Endpoint of structure formation (CHEX-MATE), and the images of the LOFAR Two-meter Sky Survey-Data release 2 (LoTSS-DR2). The cluster is undergoing a merger along the north-south axis, and shows a central radio halo and two radio relics, one in the southern and one in the northern regions. The analysis of the intracluster medium (ICM) distribution revealed the presence of a northern surface brightness (SB) jump associated to the merger event. By extracting spectra across this discontinuity, we classified the edge as a cold front. Furthermore, we made use of upgraded Giant Metrewave Radio Telescope observations that allowed us to perform a spectral analysis of the G113 radio emission. We found evidence of re-acceleration of particles in the northern relic, and we measured an associated Mach number of $\mathcal{M}=1.95\pm0.01$, as inferred from radio observations. We then performed a point-to-point analysis of the X-ray and radio emission both in the halo and in the northern relic regions. We found a strong correlation for the halo and an anti-correlation for the relic. The former behaviour is in agreement with previous studies. The relic anti-correlation is likely related to the reverse radial distribution of the X-ray (increasing towards the cluster centre) and radio (decreasing towards the cluster centre) emissions. Finally, we performed a point-to-point analysis of the radio emission and the residuals obtained by subtracting a double beta model to the X-ray emission. We found a strong correlation between the two quantities. This behaviour suggests the presence of a connection between the process responsible for the radio emission and the one that leaves fluctuations in the X-ray observations.}

\titlerunning{A combined LOFAR and XMM-Newton analysis of the disturbed cluster PSZ2G113.91-37.01}
\authorrunning{M. G. Campitiello et al.}
   \maketitle

\section{Introduction}\label{Introduction}
According to the hierarchical scenario of structure formation, clusters of galaxies form through accretion of mass and infall of smaller
sub-structures \citep[e.g.][]{Frenk1983, Kravtsov2012}. Due to the large amount of energy dissipated (up to 10$^{63}$-10$^{64}$ ergs during one cluster crossing time, that is $\sim$ Gyr), these phenomena have a strong impact both on the thermal and on the non-thermal component of clusters. The infall of matter in the cluster potential generates shock waves and cold fronts in the Intra Cluster Medium (ICM) that appear in the form of sharp surface brightness (SB) discontinuities in the X-ray observations \citep[e.g.,][for a review]{Markevitch2007}. These features, together with the merger-induced turbulence, are often involved in the formation of cluster-scale extended non-thermal structures, that are classified as radio halos or as radio relics, depending on their shape, location and radio properties \citep[e.g.,][for observational reviews]{Feretti2012,vanWeeren2019}. In particular, radio halos are usually co-located with the main X-ray emission of the hosting cluster, while radio relics are mainly located in cluster peripheries. Furthermore, these latter are characterised by an arc-like morphology and are observed to be polarised ($\sim$ 20-30\% at 1.4 GHz). 

The origin of these radio sources has not yet been fully understood. At the moment, theoretical models \citep[see][]{Bruggen2012,Brunetti2014} suggest that radio halos originate by re-acceleration mechanisms driven by merger-induced turbulence, whereas radio relics are generated by weak shocks (Mach number $\mathcal{M}$ = 2--4). Many observations seem to agree with the predictions of these models. For example, both relics and radio halos are mostly detected in merging galaxy clusters \citep[e.g.,][]{Buote2001,Bruno2021,Cassano2023}, and the former ones, with their arc-shape morphology and high polarisation, are often observed at the location of shocks \citep[e.g.,][]{akamatsu2013,Botteon2016,Eckert2016}. However, the details of this global picture are still not clear. For both relics and halos, the main issues are related to the origin of the seed particle population, the (re)acceleration mechanism and the role of the magnetic field.  
The weakness of the relic scenario, resides in the fact that, if we assume that particles are accelerated directly from the thermal pool, the shocks detected are not strong enough to reproduce the observed relic radio spectrum and luminosity \citep{Botteon2016a,Botteon2020a,Eckert2017,Hoang2017}. To overcome this limit, models assuming re-acceleration of clouds of pre-existing relativistic particles in the ICM have been proposed \citep[e.g.][]{Markevitch2005,Kang2012,Pinzke2013}. These are supported by the connection between relics and old active galactic nuclei (AGN) plasma observed in some systems \citep[e.g.,][]{Bonafede2014, vanWeeren2017}. For radio halos, the opened questions are related to the understanding of the complex mechanisms that drain a fraction of the energy of the large-scale turbulence generated by cluster formation events (such as mergers), into electromagnetic fluctuations and collisionless mechanisms of particle acceleration at much smaller scales \citep[e.g.,][]{Brunetti2011, Miniati2015, Brunetti2016, Sironi2022}. Also the definition of the role played by cosmic ray protons (CRp) and accelerated protons in the generation of these extended radio sources is still debated \citep[e.g.][]{Brunetti2017,Nishiwaki2022}.

From this introduction, it appears that combined X-ray and radio analyses of a large number of galaxy clusters are needed to deepen our knowledge on the interplay between the thermal and the non-thermal component of these systems. For example, the identification of a spatial correlation between the X-ray and radio surface brightness ($I_X-I_R$) allows to verify, and eventually to constrain, the connection between these two components \citep[as already discussed in e.g.,][]{Govoni2001,Brunetti2004,Pfrommer2008,Donnert2010,Brunetti2014}.
In this paper, we perform a combined X-ray and radio analysis of the cluster PSZ2G113.91-37.01 (hereinafter, G113). 
This cluster was first detected by the ROSAT All Sky Survey \citep[RASS,][]{Voges1999} and is included in the meta-catalogue of X-ray detected clusters of galaxies \citep[MCXC][]{Piffaretti2011} with the name of MCXC J0019.6+2517 and an associated redshift of $z=0.1353$. The more recent Planck survey re-assessed the cluster redshift to $z=0.3712$ and measured a mass of $M_{500}= (7.58 \pm 0.55)\times10^{14} M_{\odot}$ \citep{Planck2015}. Despite the ROSAT detection, no observation by any major X-ray satellite has been performed before the ones used in this work and realised by the XMM-Newton telescope as part of the ``Cluster HEritage project with XMM-Newton: Mass Assembly and Thermodynamics at the Endpoint of structure formation'' (CHEX-MATE), an XMM-Newton Multi-Year Heritage program \citep[see][for more details\footnote{http://xmm-heritage.oas.inaf.it/}]{CHEX-MATE2021}. From the radio side, new instruments able to perform a deep and a high resolution imaging have been developed only for a few years. The principal dataset used in this analysis belongs to the fairly recent LOFAR Two-meter Sky Survey-Data release 2 \citep[LoTSS, LoTSS-DR2][]{Shimwell2022}, an ongoing radio survey that aims to observe the entire northern sky in the frequency range 120 $-$ 168 MHz with the LOFAR High Band Antennas (HBA). We refer the reader to \cite{Shimwell2017,Shimwell2019} for more details on LoTSS and its scientific goals. G113 belongs to the LoTSS-DR2/PSZ2 sample presented in \cite{Botteon2022} and shows a central radio halo, and a couple of radio relics that are developing in opposite directions. The power of combining the CHEX-MATE and LoTSS data has been already tested in a couple of single cases, such as A2249 \citep[or PSZ2 G057.61+34.93,][]{Locatelli2020} and A1430 \citep[or PSZ2 G143.26+65.24,][]{Hoeft2021}. This work remarks again the power of the synergy between the two datasets, setting the stage for a full exploitation performed with the entire CHEX-MATE sample. Finally, we complemented our analysis with upgraded Giant Metrewave Radio Telescope (uGMRT) observations that allowed us to investigate the spectral radio properties of G113.

This Paper is structured as follows: in Sect. \ref{Sect:dataprep}, we summarise the radio and X-ray data reduction and the respective imaging processes; in Sect. \ref{Sect:analyses}, we describe the analysis performed to investigate the properties and correlations of the X-ray and radio emissions; we discuss the results and we draw our conclusions in Sect. \ref{Sect:Discussion}. We adopted  a flat $\Lambda$-cold dark matter cosmology with $\Omega_M(0) = 0.3$, $\Omega_{\Lambda} = 0.7$, $H_0 = 70$ km Mpc s$^{-1}$. The scale factor at the redshift of the cluster is: 5.129 kpc/".

\section{Data preparation} \label{Sect:dataprep}

\begin{figure*}
    \hspace{-1.05cm}
    \includegraphics[scale=0.51]{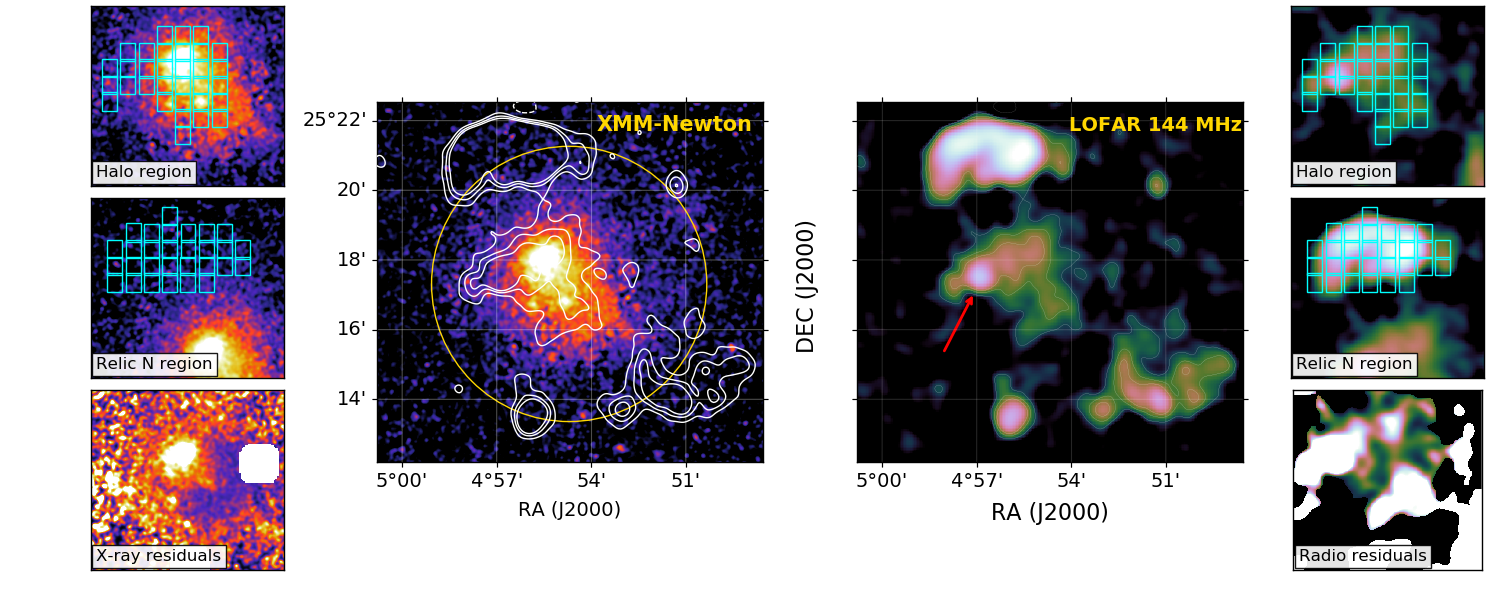}
    \caption{Central image left: G113.91 XMM-Newton observation in the 0.7–1.2 keV band with LOFAR radio contours ([-3,3,6,9,12]$\times \sigma$, with $\sigma=0.37$ mJy/beam) superimposed. The gold circular region shows the extension of $R_{500}$. Central image right: LOFAR image of G113 with the same contours used for the X-ray image. The resolution of the radio map is 33"$\times$27". The red arrow in the LOFAR panel shows the location of the ambiguous blob radio emission that was not subtracted from the halo emission. Subplots on the left (right) from top to bottom: zoom on the X-ray (radio) halo region, zoom on the X-ray (radio) northern relic region, residuals obtained by dividing the X-ray (radio) image for the best-fit double $\beta-$model (exponential model, see Subsect. \ref{ppanalysis} for more details). }
    \label{fig:G113}
\end{figure*}

\subsection{LOFAR observations}\label{Sect:dataprepradio}


LoTSS pointings are characterised by a full width at half maximum of $\ang{3.96}$ at 144 MHz and are separated by $\sim\ang{2.6}$. For this reason, a specific target is usually covered by multiple pointings. In our case, the cluster G113 falls in two pointings of the survey, namely P004+26 and P005+22
that have to be combined. In order to correct for direction-independent and direction-dependent effects that are
present in the data, pointings were processed with fully automated pipelines developed by the LOFAR Surveys Key Science Project team. The calibrated data used in this analysis are the same of \cite{Botteon2022}. For this reason, we refer the reader to that work for more details on the reprocessing steps.
To perform the imaging of our dataset, we resorted to the WSClean v2.8\footnote{https://wsclean.readthedocs.io/en/latest/index.html} algorithm \citep{Offringa2014}. As first step, we removed the contribution of discrete sources in the halo region by imaging the data sets with a $uv$ cut corresponding to a physical scale of 500 kpc at the cluster redshift. By observing the model image obtained, we noticed the presence of an ambiguous radio source in the north east part of the radio halo (see the red arrow in Fig. \ref{fig:G113}, central panel). Due to the lack of an optical counterpart, we decided to include it in our analysis, and we will discuss its treatment in Subsect. \ref{ppanalysis}. The clean components of the other discrete sources co-spatial with the halo were then subtracted from the visibility data. Sources detected in correspondence of the radio relics where not subtracted. Starting from the visibility data, we produced images at different resolutions to better observe the nature and morphology of the diffuse radio emission in the ICM. In particular, we adopted the \cite{Briggs1995} weighting scheme with \texttt{robust}=-1, -0.5 and -0.25, and we applied Gaussian $uv$ tapers in arcsec equivalent approximately to 50, and 100 kpc at the cluster redshift. Furthermore, we adopted the multi-scale multi-frequency deconvolution option \citep{Offringa2017} and subdividing the bandwidth into 6 channels for all imaging runs. By comparing the radio flux of the different images, we observed that the one that maximise it in the region of the radio halo is the image obtained with \texttt{robust}=-0.25 and \texttt{taper}=100 kpc. Therefore, we decided to use this image for our analysis. The resolution and rms of the final image are 33 arcsec $\times$ 27 arcsec and 0.37 mJy/beam, respectively.

\subsection{uGMRT observations}

To investigate the spectral nature of the G113 radio emission, we analysed observations obtained with the uGMRT under project code 40\_025. These were realised in Band 4 (550–750 MHz), with an observing time of $\sim6$ h. Data were recorded in 2048 frequency channels with integration time of 5.3 s in total intensity mode. The 200 MHz bandwidth was split into four frequency slices with a bandwidth of 50 MHz each centred at 575, 625, 675 and 725 MHz. Data were then processed independently using the Source Peeling and Atmospheric Modeling pipeline \citep[SPAM,][]{Intema2009} and setting the absolute flux density scale according to \cite{Scaife2012}. After calibration, the four slices were jointly deconvolved with WSClean. The same $uv$-cut used for the LOFAR observations was adopted in order to subtract discrete sources with a physical scale lower than 500 kpc (at the cluster redshift) in the halo region. The imaging of these output calibrated data will be described in Subsect. \ref{Sect:radioem}. 

For both the LOFAR and uGMRT images, the uncertainties on the flux estimates were computed following the equation:
\begin{equation}
    \Delta S = \sqrt{(\sigma \cdot \sqrt{N_{beam}})^2+(\xi_{cal} \cdot S)^2 + (\xi_{sub} \cdot S)^2 }
\end{equation}
with $\sigma$ being the RMS noise of the map, N$_{beam}$ the number of independent beams within the considered region, $\xi_{cal}$ the calibration error, and $\xi_{sub}$ the error associated to the source subtraction. This latter was computed by estimating the ratio between the flux density measured before and after the subtraction in the regions where discrete point sources were detected and subtracted. It was included only in the estimation of the halo emission errors, and not in the relics one. No subtraction was indeed applied to these latter diffuse radio sources. We adopted $\xi_{cal}$ = 10\%  and $\xi_{sub}$ = 4\% for LOFAR, and $\xi_{cal}$ = 5\%  and $\xi_{sub}$ = 3\% for uGMRT images.

\subsection{XMM-Newton observations}\label{Sect:dataprepX}
The observation data file (ObsID: 0827021001) was downloaded from the XMM-Newton archive and processed with the XMMSAS \citep{Gabriel2004} v18.0.0 software for data reduction. The production of calibrated event files from raw data was realised by running the tasks \texttt{emchain} and \texttt{epchain}. We only considered single, double, triple, and quadruple events for MOS (i.e., PATTERN $\leq$ 12) and single events for pn (i.e., PATTERN $\leq$ 4), and we applied the standard procedures for bright
pixels and hot column removal (i.e., FLAG==0), and pn out-of-time correction. The background subtraction was performed by removing step-by-step all the individual background components. First, we removed the photon X-ray background (i.e., unresolved cosmological sources plus the Local Bubble and the Galactic Halo). The contribution of this component was estimated by fitting simultaneously the {\it XMM-Newton} data within $R_{500}$ with the {\it ROSAT} All-Sky Survey spectrum extracted from the region just beyond the virial radius using the available tool\footnote{http://heasarc.gsfc.nasa.gov/cgi-bin/Tools/xraybg/xraybg.pl} at the
HEASARC webpage. Then, we produced an image of this components, based on the predicted count-rate per pixel and accounting for the XMM-Newton vignetting as derived from  the calibration files, and we subtracted it from the raw image. Second, we removed the contribution of the quiescent particle background by subtracting the image obtained from the renormalised Filter Wheel Closed observations following the procedure explained in \cite{Zhang2009}. Finally, we removed the contribution of the residual soft protons by producing an image of this components based on the expected count-rate per pixel, which was determined by adding a power-law component (with both the slope and normalisation free to vary), folded only with the Redistribution Matrix File, to the
background modelling of the spectra within $R_{500}$. The net exposure time obtained after the cleaning is: 54.2 ks, 53.5 and 49.4 ks for the mos1, the mos2 and the pn, respectively. To detect the point-like sources we first used the \texttt{edetect-chain} task. Then we visually checked the result. We noticed that a bright source located in the southern part of the core was not identified by the task. We decided to include it in our analysis and we investigate and discuss its nature in Sect.~\ref{Sect:ICM}.

The final images used in this work are filtered in the [$0.7,1.2$] keV energy band and are characterised by a binning of 50 physical pixels, which corresponds to a pixel size of 2.5 arcsec. The holes related to point-like source subtraction were filled with the \texttt{dmfilth} CIAO task.


\section{X-ray and radio analyses} \label{Sect:analyses}

\subsection{Analysis of the ICM distribution}\label{Sect:ICM}
\begin{figure*}
    \centering
    \includegraphics[scale=0.7]{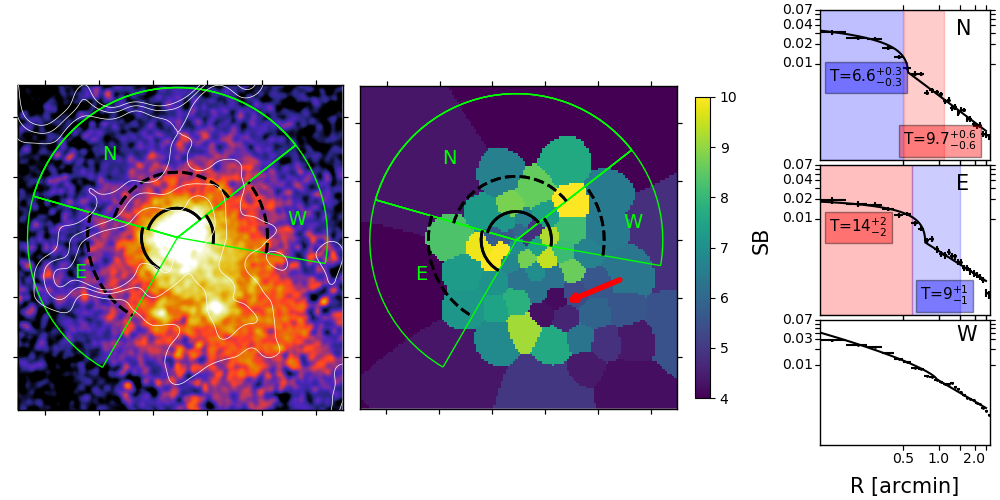}
    \caption{Left: X-ray image of G113.91. The green regions represent the three sectors adopted for the extraction of the SB profiles. The black solid lines represent the position of the edge and the boundary of the inner sector where the temperature was measured. The black dashed lines instead represent the boundary of the outer sector where the temperature was measured. Center: 2D temperature map of G113. The colorbar is in keV units, and the red arrow show the position of the cold region detected in the cluster. Right (from top to bottom): SB profiles extracted in the northern (N), eastern (E), western (W) sectors. The black continuous lines represent the best fit of the eastern ($\chi^2/ d.o.f=0.801$ and $C=1.94\pm0.17$), northern ($\chi^2/ d.o.f=1.56$ and $C=1.9\pm0.2$) and western ($\chi^2/ d.o.f=7.02$ and $C=1.14\pm0.04$) jump. The temperatures reported in the blue and red boxes are in keV units and are the temperature measured before and after the SB discontinuities. The corresponding blue and red shaded area show the regions where the temperatures were measured.}
    \label{fig:jump_temperature}
\end{figure*}
According to the morphological classification performed in \cite{Campitiello2022}, G113 is one of the most disturbed systems of the CHEX-MATE sample, with concentration $c=0.23^{+0.03}_{-0.04}$, centroid shift $w=0.34^{+0.08}_{-0.09} 10^{-1}$ and $M=0.74$. This latter quantity ($M$) is a combination of $c$, $w$ and of the power ratios of order $P_{2}/P_0$ and $P_{3}/P_0$. We refer the reader to \cite{Campitiello2022} for more details on the definition of all these parameters. The XMM-Newton image of the cluster confirms that this system is undergoing a merger along the N-S axis, and from a first visual check stands out the presence of an edge in the X-ray surface brightness (SB) distribution located in the northern direction. We thus extracted a profile across this region using the northern sector N shown in Fig. \ref{fig:jump_temperature} (left panel).  The software used for this step of the analysis was the python package PYPROFFIT \citep[see][, for more details]{Eckert2020}.
 The SB profile of this discontinuity was then modelled assuming that the underlying density profile follows a broken power law \citep[e.g.,][and references therein]{Markevitch2007}, and that, due to the energy band of our observations, the temperature and the abundance are uniform. In the case of spherical symmetry, the downstream and upstream (subscripts $d$ and $u$) densities differ by a compression factor $\mathcal{C}=n_d/n_u$ at the distance of the jump $r_j$:

\begin{equation}
\begin{aligned}
n_d(r) &= \mathcal{C} n_0 \bigg( \frac{r}{r_j}\bigg)^{\alpha_1}  \, \, \, \text{if} \,\,\,  r\leq r_j  \\
n_u(r) &= n_0 \bigg( \frac{r}{r_j}\bigg)^{\alpha_2}  \, \, \, \, \,\,\, \text{if} \,\,\,  r> r_j,
\end{aligned}
\end{equation}
where $\alpha_1$ and $\alpha_2$ are the power-law indices, $n_0$ is a normalisation factor, and $r$ denotes the radius from the centre of the sector. The centre of the sector was chosen in order to optimise the detection of the edge. All these quantities were free to vary during the fitting procedure. 
Our result confirms the presence of a jump in the northern region of the cluster, with $\mathcal{C}=1.9\pm0.2$ ($\chi^2/ d.o.f=1.56$). The position and angular extension of this discontinuity is shown in Fig. \ref{fig:jump_temperature} (left panel, solid black line in the N sector). In order to investigate the nature of the jump we first realise a  2D temperature map of the cluster using the Voronoi tessellation method. To define the Voronoi map bins, we considered the WVT algorithm provided by \cite{Diehl2006}, which is a generalisation of \cite{Cappellari2003} Voronoi binning algorithm. We required a signal-to-noise $S/N \sim 30$, which allowed us to estimate the temperature with a statistical uncertainty of 10-20 \% \citep[see][for more details]{Lovisari2019}. We then measured the projected temperature in each region of the map. The final result is shown in Fig. \ref{fig:jump_temperature} (central panel), while the associated error map is shown in the Appendix \ref{Appendix:errormap}, Fig. \ref{fig:errmap} (right panel). We found that a cold region $T \sim 5$ keV is visible in the northern part of the cluster. However, due to the coarse Voronoi tassellation it is not possible to clearly understand whether it corresponds or not to our SB discontinuity. Thus, we extract a spectra from two annuli with $S/N \sim 50$, selected with the same aperture and centre used for the extraction of the SB profile, and covering the region before and after the discontinuity. We then fitted the spectra with an APEC thermal plasma model with an absorption fixed at $n_H= (2.96 \pm 0.08) \cdot 10^{20}$ cm$^{-2}$ (Bourdin et al. in prep), and then measured the
projected temperature in each region of the map.
We found a temperature of $T=6.6\pm0.3$ keV in the inner sector and a temperature of $T=9.7 \pm 0.6$ keV in the outer one. 
In order to understand whether this feature is due to the presence of a cool core and it is thus a general property of the whole profile, we extracted SB profiles also from the eastern (E) and western (W) regions. We excluded the southern region due to its extended tail of X-ray emission. We found that another jump, of magnitude $\mathcal{C} = 1.94 \pm 0.17$ ($\chi^2/d.o.f = 0.801$) is present in the E direction. To investigate its nature, we extracted a temperature profile using the same procedure described for the N sector. We found an inner temperature of $T=14\pm2$ keV and an outer temperature of $T=9 \pm 1$ keV. In the western region instead, no SB discontinuity was detected. In particular, by fitting the SB profile we found a jump of magnitude $\mathcal{C}= 1.14 \pm 0.04$ and a $\chi^2/d.o.f = 7.04$. This latter parameter shows that this model is not able to properly fit the profile. Also in this case we measured the temperature using the same radial range adopted for the eastern discontinuity. We found an inner temperature of $T=6.1 \pm 0.4$ keV and an outer temperature of $T=8 \pm 1$ keV. The results arising from the analysis of these three sectors prove that the edge detected in the northern region is not tracing an overall property of the core. We thus classified the northern edge as a cold front. For what concerns the eastern discontinuity, we noticed that the temperatures measured in the eastern sectors are higher than the ones measured in the western ones. Furthermore, the eastern sector used for the extraction of the SB and temperature profiles covers the region where the ambiguous radio source presented in Sect. \ref{Sect:dataprepradio} (row arrow in Fig. \ref{fig:G113}) is located. The temperature gradient observed, with high temperature in the inner annulus and lower temperature in the outer one, may be either due to the presence of an AGN or due to the accretion of a small sub-clump of galaxies which generates a shock. We estimated the Mach Number that, if confirmed, this shock would have  Using the Rankine–Hugoniot conditions \citep[see, e.g.][]{Markevitch2007}, we can link the density and temperature jump to the Mach number through the relations:
\begin{equation}\label{eq:densityjump}
    \mathcal{M}^2 = \frac{3\mathcal{C}}{4-\mathcal{C}}
\end{equation}
and 
\begin{equation}\label{eq:temperature_jump}
    \mathcal{M}^2 = \frac{(8T_{j}-7)+[(8T_{j}-7)^2+15]^{1/2})}{5}
\end{equation}
where $T_j$ is the ratio between the temperature post and pre-shock. From eq. \ref{eq:densityjump} we found $\mathcal{M} = 1.68\pm 0.14$, while from eq. \ref{eq:temperature_jump} we found $\mathcal{M}=1.55\pm 0.04$. Considering the uncertainties, these two estimates are in agreement. However, the hypothesis provided are just hints for possible interpretation: further investigations are needed to understand whether the radio and X-ray emission are linked and to derive any conclusion.

The temperature map revealed also the presence of another cold region, located in the southern part of the cluster and characterised by a temperature $T \sim 4$ keV. This corresponds to an X-ray clump that was not identified as a point source during the imaging of the XMM-Newton images (see Sect.~\ref{Sect:dataprepX}). By consulting optical images of the DESI Legacy Imaging Surveys\footnote{https://www.legacysurvey.org/}, we noticed the presence of two foreground galaxies located at redshift $z=0.133$. This may explain the original wrong RASS estimation of the cluster redshift, that was assessed to be $z=0.1353$ instead of $z=0.370$ (see Sect. \ref{Introduction}).

\subsection{Analysis of the diffuse radio emission}\label{Sect:radioem}
\begin{figure*}
    \centering
    \includegraphics[scale=0.75]{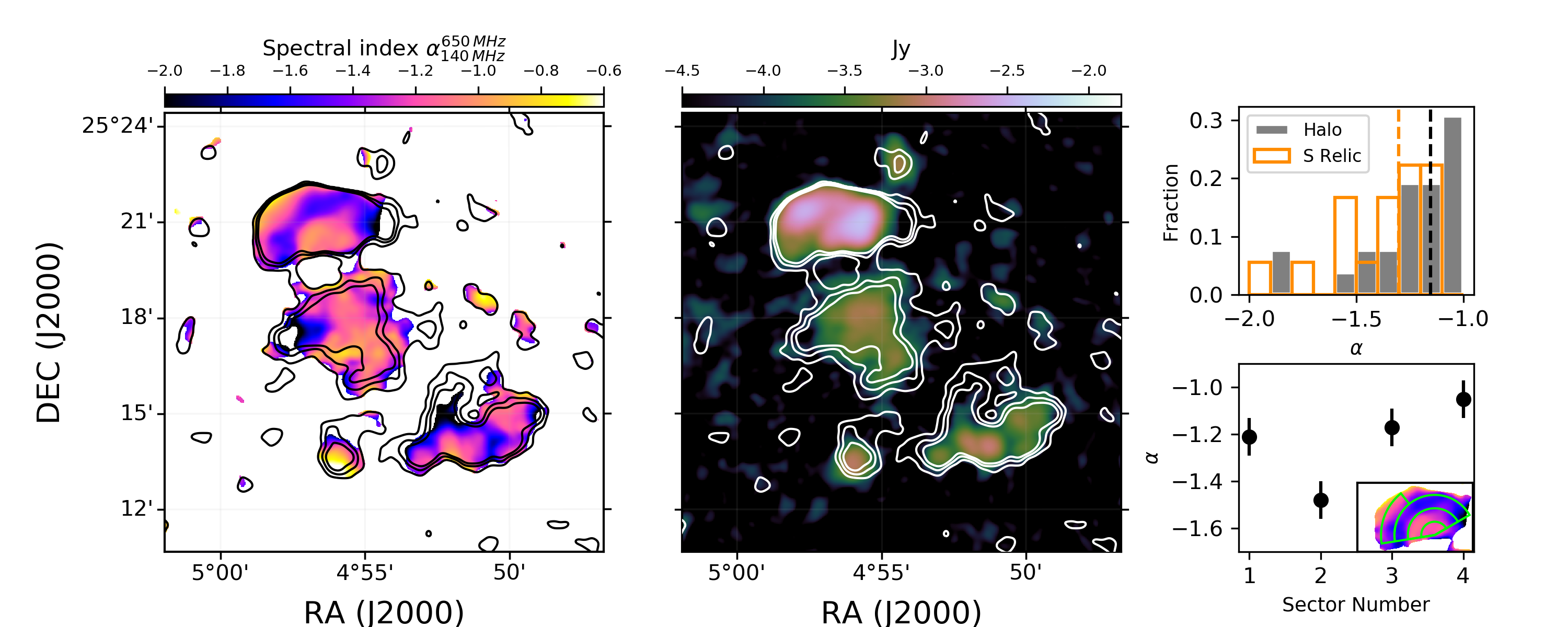}
    \caption{Left: spectral index map of the cluster G113 with LOFAR contours of the image at 35"$\times$35" overplotted ([-3,3,6,9,12]$\times \sigma$, with $\sigma$ = 0.37 mJy/beam). Centre: uGMRT image of G113, with LOFAR contours overplotted. The resolution of the image is 35"$\times$35" and the rms noise is $\sigma=0.037$ mJy/beam. Right top: Distribution of the spectral index extracted from square-shaped boxes of width 35" in the halo (grey) and southern relic (orange) regions. The vertical lines represent the $\alpha$ median values for the halo (black) and relic (orange). Right bottom: values of $\alpha$ extracted from the sector shown in the subplot. The sectors are numbered from the outer annulus (1) to the inner one (4). }
    \label{fig:spectral_index}
\end{figure*}

From a radio point of view, G113 represents a rare case of a cluster hosting both a central radio halo and a pair of radio relics developing one in the northern and one in the southern regions of the cluster. The axis on which these two latter structures lie, traces the direction of the merger event that is affecting the cluster and that is visible in the X-ray images (see Subsect. \ref{Sect:ICM}).

\begin{figure}
    \centering
    \includegraphics[scale=0.5]{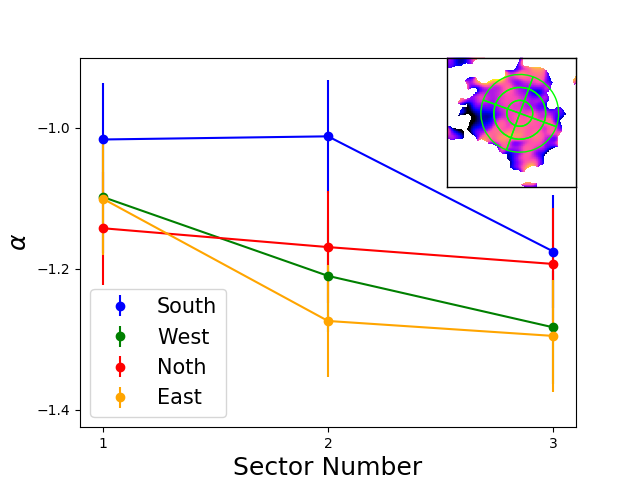}
    \caption{Radial profile of $\alpha$ in the radio region. The four profiles refer to four different sectors that cover the northern, southern, eastern and weastern regions, as shown in the cluster subplot.}
    \label{fig:settori}
\end{figure}

In order to investigate the properties of the non thermal emission of G113, we realised a spectral index map. To achieve this goal, we produced LOFAR and uGMRT images with a common inner $uv$-cut to compensate for the different interferometer $uv$-coverage, which is critical for the recovery of the extended emission \citep[e.g.,][]{Bruno2023}. Furthermore, we adopted the same input parameters (multiscale, robust and taper) presented in Subsect. \ref{Sect:dataprepradio}, and we fixed the shape of the restoring beam in order to have a final resolution of 35"$\times$35" in both images. We then re-gridded the two images to the same pixel grid (i.e. the LOFAR one). The RMS noise of the maps is $\sigma = 0.30$ and $\sigma = 0.037$ mJy/beam for LOFAR and uGMRT images, respectively, and pixels with flux values lower than 2$\sigma$ were masked for each frequency. The spectral index map was then obtained by computing for each pixel the alpha index, defined as $\alpha^{144}_{650} = \log(S_{650}/S_{144})/\log(\nu_{650}/\nu_{144})$ where $S$ is the flux density and $\nu$ the considered frequency. The associated uncertainties were obtained using the relation:
\begin{equation}
    \Delta \alpha = \frac{1}{\ln(\frac{\nu_{650}}{\nu_{144}})}\sqrt{\bigg(\frac{\Delta S_{650}}{S_{650}}\bigg)^2+\bigg(\frac{\Delta S_{144}}{S_{144}}\bigg)^2}
\end{equation}
The final spectral index map is shown in Fig. \ref{fig:spectral_index} (left panel), while the associated error map is reported in the Appendix \ref{Appendix:errormap}, Fig. \ref{fig:errmap} (left panel). It is possible to observe that the spectral index measured in the halo region, varies between $-0.8$ and $-2$. We also noticed that higher values of $\alpha$ (i.e., $\alpha \geq -1$) seems to be mainly found in the southern and in the north-west regions of the halo. We thus extracted a profile of $\alpha$ across four sectors that cover the northern, southern, western and eastern region of the radio halo. The result, together with the sectors adopted are shown in Fig. \ref{fig:settori}, and the associated uncertainties are 1 $\sigma$ errors. By looking at the central value measured in each annulus of the considered sectors, it seems that a steepening of $\alpha$ is present. Furthermore, the spectral index measured in the northern and southern regions seems to follow a different trend. However, the uncertainties of the map do not allow us to clearly confirm this behaviour. 
However, the location of the X-ray discontinuity (close to the northern boundary of the radio halo) does not allow to obtain an accurate estimate of the variation of the $\alpha$ index across the cold front. In the bottom plot of Fig. \ref{fig:spectral_index} (right panel) we report the spectral index values extracted from square-shaped boxes with width 35" (that is the beam width of the map) covering the halo region. We estimated a mean $\alpha$ value of -1.2, and an associated standard deviation of $\sigma_{\alpha}=0.2$. This value is more than two times higher than the median $\alpha$ error, $\alpha_{err} \sim 0.09$, measured in the whole halo region, thus suggesting that the observed variations are not (only) related to systematic errors, but are linked to the presence of intrinsic small-scale fluctuations in the spectral index across the halo. A similar behaviour was already observed in e.g., \cite{Botteon2020a}, \cite{Kamlesh2023}, but not observed in others \citep[e.g.,][]{vanWeeren2016a}. For what concerns the northern relic, we observed an evident steepening of the spectral index towards the cluster centre. To properly characterise this behaviour, we computed the average value of $\alpha$ in the four concentric rings shown in the cluster subplot of Fig. \ref{fig:spectral_index} (bottom right panel). Moving from north to south, the trend observed consists in a first decrease of $\alpha$ (from the outer annulus marked as 1 to following one), followed by a subsequent increase (from the annulus 2 to the inner annuli 3 and 4 ). By assuming that the particles are being accelerated via dynamic site acceleration (DSA) by a shock that is moving outwards, we can use the spectral index measured in the outer annulus (1) to estimate the Mach number, $\mathcal{M}$ \citep{Blandford1987}. The relation that links $\alpha$ to $\mathcal{M}$ is:
\begin{equation}
\mathcal{M} = \sqrt{\frac{2\alpha_{inj}-3}{2\alpha_{inj}+1}}
\end{equation}
Starting from $\alpha_{inj}=-1.21\pm0.07$ measured in the outer annulus, we found $\mathcal{M}= 1.95 \pm 0.01$. The $\alpha$ steepening observed between the annuli 1 and 2 can be explained with synchrotron ageing. The subsequent increase of $\alpha$ may be instead related to the presence of a radio source that is injecting electrons in the surrounding regions, and that is visible in images with higher resolution (see Fig. \ref{fig:pointosurcemap} in Appendix \ref{Appendix:pointsource}). By looking for an optical counterpart, we verified that this source corresponds to a galaxy located at the cluster redshift ($z=0.315$).
Finally, no peculiar trend of the spectral index is found in the southern relic. We thus computed the median $\alpha$ and standard deviation values using the same approach adopted for the radio halo and founding $\alpha = -1.3$ and $\sigma_{\alpha}=0.2$. Also in this case the distribution of the spectral index measured in the squared-shape boxes is shown in the histogram of Fig. \ref{fig:spectral_index} (top right panel).

\subsection{Point to point correlations}\label{ppanalysis}

Looking for the presence of a spatial correlation between the X-ray and radio emissions is a powerful tool to investigate the relation between the thermal and non-thermal cluster components \citep{Brunetti2014}. As it is possible to observe from Fig. \ref{fig:G113}, the thermal and non-thermal emission of the radio halo are basically overlapped. Many studies based on the point-to-point analysis of these two components have shown that usually the X-ray and radio flux density of the halo region follow a relation of the type $I_R\propto I_X^k$, with $k<1$ \citep[e.g.,][]{Feretti2001,Govoni2001,Giacintucci2005,Botteon2020,Bruno2021}. 

\begin{figure*}
    \centering
    \includegraphics[scale=0.6]{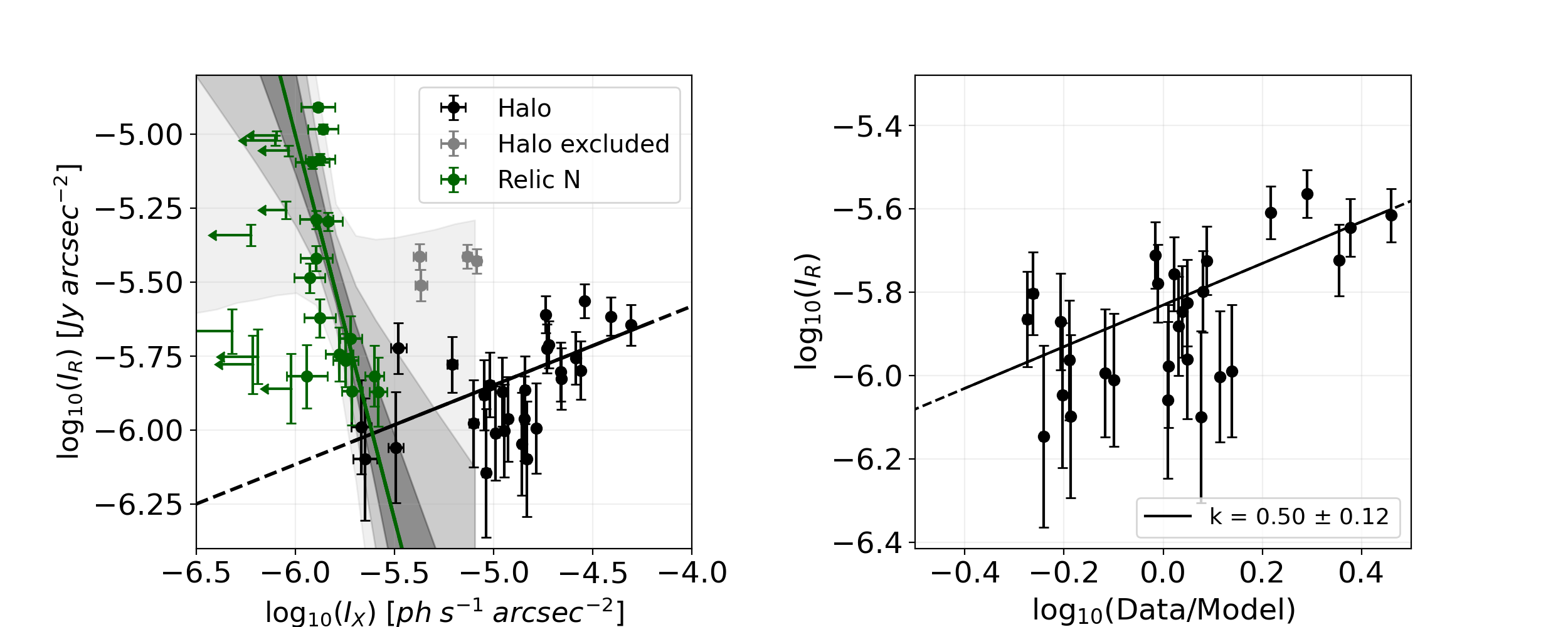}
    \caption{Left: Correlations between $I_R$ and $I_X$ computed in the halo (black) and northern (green) relic regions using the meshes shown in the subplots of Fig. \ref{fig:G113}. Grey points correspond to the cells excluded from the halo analysis (see Sect.~\ref{ppanalysis}). The slopes and Spearman and Pearson coefficients measured are reported in Table \ref{Tab:corr}. The shadowed regions represent the scatter of the relic correlation (from dark to light grey, 1,2,3$\sigma$). Right: correlation between the radio emission and the X-ray residuals.}
    \label{fig:correlazioni}
\end{figure*}

In order to investigate this behaviour in G113, we performed a point-to-point analysis using the python code \texttt{PT-REX} described in \cite{Ignesti2022}. Based on a given threshold in surface brightness, the code generates a sampling mesh, whose cells reproduce the dimensions of the beam. For each cell, the values of $I_R$ and $I_X$ and their associated errors are computed. Finally, data are fitted with a power law and the Pearson ($\rho_p$) and Spearman ($\rho_s$) coefficients are estimated. We applied this procedure to the G113 halo region and we found that, if all the cells of the grid with a radio signal higher than 2$\sigma$ are considered, no strong correlation is observed ($\rho_p = 0.17 $ and $\rho_s = 0.22$). However, by observing the plot in Fig. \ref{fig:correlazioni} (left panel), it is possible to notice that four points (shown in grey in the plot) deviate from the general trend. These four points correspond to the position of the radio blob of emission that was not subtracted to the halo emission during the imaging process (see Subsect. \ref{Sect:dataprep}, for more details). This clump has no optical counterpart and ma be a remnant plasma in the ICM, or a substructure of the radio halo. To avoid contamination, we excluded these cells from the analysis and we recomputed the slope $k$ and the Spearman and Pearson coefficients. The final values are reported in Table \ref{Tab:corr}. We found a good correlation between $I_R$ and $I_X$. The $k$ value measured is lower than the one found in other clusters. However, we have to consider that, by defining the grid to adopt for the X-ray - radio comparison, we selected only the regions where the radio signal is higher than 2 $\sigma$. This threshold has been proved to reduce the impact that artefacts in interferometric radio images arising from the data processing have on the flux density estimates, especially for extended sources with low-SNR \citep{Botteon2020}. This radio-based threshold together with the four cells excluded due to the presence of a blob of emission, do not allow us to properly cover the entire G113 X-ray emission: as it is possible to see from Fig. \ref{fig:G113}, part of the west and south west emission is not considered in the analysis. This may produce a flattening of the correlation since cells with low X-ray emission are not considered. By repeating the analysis with a radio threshold set to 1 $\sigma$, we obtain larger uncertainties on the radio flux estimate, but we are able to retrieve a steeper slope, with $k = 0.41\pm 0.08$. This behaviour is in agreement with \cite{Bonafede2022}, where it was found that the correlation $I_{R} \propto I_X^k$ is steeper in the external regions of the cluster.

We then tested this method also on the northern and southern relics in order to understand whether correlation between the X-ray and radio emissions are present also in these diffuse radio sources. A similar analysis was already performed for the radio relic detected in the cluster MACS J1149.5+2223 \citep{Bruno2021}. The results shows the absence of any correlation. However, the authors highlighted that the the radio relic shows very peculiar features, such as an orientation of its major axis parallel (and not perpendicular) to the merger axis. For this reason, they concluded that further investigations were needed to understand if that radio source can be really be classified as a radio relic. The meshes identified by the code for the S and N relics are shown in the respective subplots of Fig. \ref{fig:G113}, and cover the radio signal over 3$\sigma$. Since these radio sources are located in the outer regions of the cluster, we assess the significance of the X-ray signal, by measuring the signal to noise ratio in each cell, and by selecting only the ones over a specific threshold. By defining the source, background and total counts as  $C_s$, $C_b$ and $C_t = C_s +C_b$, and assuming a Gaussian error propagated in quadrature, 
the S/N ratio is:
\begin{equation}
    S/N = \frac{C_t -C_b}{\sqrt{C_t + C_b}} = \frac{C_s}{\sqrt{C_s + 2 \times C_b}}.
\end{equation}
For the northern relic we selected only the cells with $S/N>3$. For the southern relic instead, it was not possible to satisfy this condition, since a low number of cells were included. We thus decided to exclude the treatment of this diffuse radio source from our work. By considering the selected cells of the northern relic, we found that the X-ray and radio emissions are linked by a strong anti-correlation between the X-ray and radio emission. We reported the $k$, $\rho_p$ and $\rho_s$ values in Table \ref{Tab:corr}, and we show the correlation and the 1, 2, 3$\sigma$ scatter in Fig. \ref{fig:correlazioni} (left panel). We also report the position of the X-ray limits measured in the cells with $S/N<3$, and we observed that few of them are not included in the 3$\sigma$ scatter. By including the central values of these points in the fit, we estimated a slope of $k=-1.3 \pm 0.7$, which is still consistent, within the uncertainties, with the one found above.


\begin{table}[]
\centering
\caption{Results of the point-to-point analysis obtained for the halo and the northern relic. From left to right: slope of the correlation $I_R\propto I_X^k$, Pearson and Spearman coefficients. In the last row are reported the values of the correlation between $I_R$ and the X-ray residuals (shown in the right plot of Fig. \ref{fig:correlazioni}).}\label{Tab:corr}
\renewcommand{\arraystretch}{1.2}
\begin{tabular}{lccc}
\hline\hline
                      & $k$ & $\rho_p$  & $\rho_s$  \\ \hline\hline
 Halo & 0.27$\pm$0.09  & 0.56 & 0.61  \\
Relic N & -2.6 $\pm$ 0.9  & -0.67  & -0.54  \\
Halo ($I_R\propto I_{X,res}^k$) & 0.50 $\pm$ 0.12 & 0.62 & 0.51 \\ \hline
\end{tabular}

\end{table}


\subsection{Residuals of the radio and X-ray emissions}
We exploit the point-to-point analysis to investigate also the presence or absence of correlations between the residual emission obtained by dividing the X-ray and radio observations by a model emission of the thermal and non-thermal component. This analysis allow us to explore possible links between the processes that leaves fluctuations in the X-ray and radio distributions. 

For what concerns the X-ray side, a double $\beta-$model is considered a good representation of the SB emission of those galaxy clusters that are characterised by a relaxed dynamical state. On the contrary, clusters that have been undergone merger events, AGN outbursts or sloshing are expected to show deviations from this smooth model. By dividing the X-ray observation of G113 for the best fit model of its SB emission, we thus expect to highlight features generated by the ongoing merger event, such as substructures and discontinuities associated to shocks and cold fronts. To obtain the residuals of the X-ray emission of G113 we proceeded as follows. We resorted to the python package \texttt{PYPROFFIT} \citep[see][for more details]{Eckert2020} to extract an averaged radial profile of the cluster. The annuli adopted for the extraction are centred on the centroid of the cluster, cover a region included in $R_{500}$ and have an elliptical shape. The centroid of the cluster, the axis ratio (major/minor) and the position angle of the elliptical annuli were calculated with the principal component analysis (PCA). This method is usually used to reduce the number of dimensions needed to describe a dataset. In our specific case, it allows us to identify the two axis that hold the highest amount of variance, and that better represent the SB distribution of the cluster (that are the axis of the elliptical annuli). The ellipse axis-ratio and position angle (from the Right Ascension axis, R.A.) found are respectively 1.26 and 129$^{\circ}$. We fit the radial profile with a double $\beta$-model, using the definition:
\begin{equation}
    I(r)=I_0 \bigg [\bigg ( 1+ \bigg (\frac{x}{r_{c,1}}\bigg )^2 \bigg ) ^{-3\beta + 0.5} + R \bigg ( 1+ \bigg (\frac{x}{r_{c,2}}\bigg )^2 \bigg ) ^{-3\beta + 0.5} \bigg]+B \text{,}
\end{equation}
and finding $\beta = 1.44^{+0.13}_{-0.11}$, $r_{c,1} = 1.94^{+0.14}_{-0.11}$, $r_{c,2} = 1.73^{+0.12}_{-0.11}$ and $R = 0.90^{+0.04}_{-0.03}$. The high $\chi^2/d.o.f. = 42.5/14$ and the unusually high $\beta$ confirm the extremely disturbed X-ray morphology of this system. Starting from this model, we created a vignetting-corrected model image representing the expected emission of the cluster. Finally, we divided the original X-ray image for the model, in order to obtain a residual map that highlights the regions were an excess or a deficiency in SB is present with respect to the model. Since both the deviations from a double $\beta$-model and the radio emission of the halo are originated by the ongoing merger event, we investigated whether correlations are present between the X-ray residuals and the radio flux. To perform a point-to-point correlation between these two quantities, we considered the same grid produced by the \texttt{PT-REX} algorithm (see subplots of Fig.~\ref{fig:G113}, which was already adopted in Sect.~\ref{ppanalysis}). 
Also in this case, we excluded the cells corresponding to the ambiguous radio blob in the halo region. For each cell included in the grid, we measured the radio flux and the X-ray residual emissions. We found that, a good correlation between these two quantities (see Fig. \ref{fig:correlazioni}, right plot), with $k = 0.50 \pm 0.12$, $\rho_p = 0.62$ and $\rho_s = 0.51$. This result depends on the model used for the description of the X-ray emission. However, a double $\beta$ model centred on the centroid (which represents the centre of the potential well) of the system is expected to be one the best representation possible to describe the SB distribution of the cluster emission. Furthermore, the plot seems to show two regimes: $\log \text{(data/model)} < 0.2$, where $I_R$ varies in a fixed range, and $\log\text{(data/model)} > 0.2$, where $I_R$ varies in another, higher range. We investigated whether the five points with $\log\text{(data/model)} > 0.2$ are either spread out or related to a specific region. We found that four of the points correspond to four of the cells that cover the X-ray core. In Fig. \ref{fig:redcells}, we show in red the position of these cells.

Historically, radio halos were classified as "smooth" sources with "regular" morphologies \citep[see][for the definition of radio halo]{Feretti2012, vanWeeren2019}. Starting from this classification, it is usual in literature to fit the radio emission of these sources with an exponential profile defined as \citep{Murgia2009}:
\begin{equation}
\label{eq:SB}
    I(r) = I_0 e^{-r/r_e}
\end{equation}
where $I_0$ and $r_e$ are the central surface brightness and the e-folding radius, respectively. However, the new generation of radiotelescopes, provided us with the opportunity to study these extended radio sources with an unprecedented resolution. Recent studies highlighted that, by increasing the S/N of the observations, it is possible to highlight an increasing number of deviations from this model \citep{Botteon2022}. Starting from our radio observations, we decided to investigate whether similar deviations are visible also in G113.
The presented fitting of Eq. \ref{eq:SB} and halo radio flux density estimations were done with a newly developed algorithm\footnote{https://github.com/JortBox/Halo-FDCA}, described in detail by \citet{Boxelaar2021}. According to this procedure, the profiles are fitted to a two-dimensional image. This code allows to fit both circular and non-circular models. By comparing the results arising from a circular and an elliptical model, we did not observe any significant difference, with the elliptical model returning almost equal semi-axes values. Given the roundish morphology confirmed by the elliptical model and given the relatively faint emission of the halo, we decided to use a circular model. 
The best-fit estimates for the peak surface brightness and radius are found through Bayesian inference and maximum likelihood estimation. To sample the likelihood function, the Monte Carlo Markov chain implemented within the emcee module is used \citep{Foreman-Mackey2013}. This method allows us to find the full posterior distribution for the model parameters. An estimation of these parameters was already presented for G113 in \cite{Botteon2022}. However, due to the application of a different uv-cut (250 kpc in \cite{Botteon2022}, and 500 kpc in this work), the point-sources subtraction may be slightly different between the two analyses and it can results in a different flux estimate. For this reason, we decided to re-perform the fit. By applying this algorithm to our cluster, we found a lower $I_0= 3.3^{+0.2}_{-0.2}\, \mu Jy/arcsec^2$ with respect to the one found in \cite{Botteon2022} ($I_0= 5.1^{+0.2}_{-0.2}\, \mu Jy/arcsec^2$ ). This confirms that our uv-cut leads to the subtraction of more components. For what concerns the e-folding radius, a good agreement was found with $r_e = 350 \pm 20$ kpc in our work and $r_e = 370 \pm 15$ kpc in \cite{Botteon2022}. Starting from these fit results we derived a model image of the radio emission. We then obtained the radio residual image by dividing the original LOFAR image for the best-fit exponential model. Both the radio and X-ray residuals are shown in the respective subplots of Fig. \ref{fig:G113}.

As in X-rays, deviations from a smooth surface brightness models may highlight the presence of substructure or discontinuities in the radio emission. Recently, \citet{Botteon2023} claimed the presence of numerous radio surface brightness discontinuities, similar to those detected in X-rays, in a number of radio halos observed with MeerKAT. Some of these radio edges were found to be associated with shocks and cold fronts, suggesting a tight connection between the thermal and non-thermal components in the ICM. As merger events disturb the cluster atmosphere and dissipate energy into the acceleration of relativistic particles, we tested whether the X-ray and radio residual emissions are linked by performing a point-to-point analysis on the residual images. Also in this case the meshes adopted are the same used in the previous analysis of the halo region. We did not find any correlation between the radio and X-ray residual emissions. This behaviour could be related to different reasons. As already shown in literature, many radio discontinuities do not show an X-ray counterpart \citep[e.g.,][]{Botteon2023}. In these cases the deviations from the exponential model may highlight specific regions in the ICM that are characterised by a high fraction of seed relativistic electrons or by a local enhancement of the acceleration efficiency or of the magnetic field strength. For these features, the absence of a correlation between the X-ray and radio residuals may be an indication of the fact that this two type of emission highlight different type of discontinuities. On the other hand, it was suggested that the S/N ratio of the observations may has a strong impact on the detection of radio edges. For example, all the radio discontinuities detected in \cite{Botteon2023}, were found in clusters that own observations with a high S/N ratio. Given the low S/N ratio of our observations, the lack of correlation between the X-ray and radio residuals may be related to the impossibility of detecting all the features, especially the ones that may be present in correspondence of the X-ray substructures. Finally, we do not found any correlation between the X-ray residuals and the $\alpha$ index of the spectral index map.


\section{Conclusions} \label{Sect:Discussion}
In this Paper, we performed a joint X-ray and radio analysis of the cluster G113, 
a merging system showing a peculiar radio morphology: besides a central radio halo, this cluster hosts also two relics perpendicular to the merger axis, located one in the northern and one in the southern region. The double relic configuration is exploited in literature to provide good constraints on cluster merger scenarios. 
Currently, a dozen objects showing this feature are known \citep{vanWeeren2019}, among which some well-studied cases are: Abell 3667 \citep{Rottgering1997}, Abell 3376 \citep{Bagchi2006}, which are respectively the first and second double relics detected, and MACS J1752.0+4440 \citep{vanWeeren2012,Bonafede2012} and Abell 1240 \citep{Kempner2001,Bonafede2009}.

The X-ray analysis of the ICM of G113 allowed us to identify a surface brightness discontinuity in the northern region of the cluster. By performing a spectra extraction from two annuli across the edge, we measured a e downstream (i.e. post-front) temperature of $T=6.6\pm0.3$ keV and an upstream (i.e. pre-front) temperature $T=9.7\pm0.6$ keV. We thus classified the discontinuity as a cold front.

In order to investigate the non-thermal properties of G113 we realised a spectral index map. We found that the halo region is characterised by a mean $\alpha$ value of -1.15, and an associated standard deviation of $\sigma_\alpha$ = 0.23. This latter value is larger than the median error and suggests that the observed variations are linked to the presence of intrinsic small-scale fluctuations in the spectral index across the halo. We also observed a flattening of the profile in the northern front of the northern relic. This behaviour could be explained by assuming that particles are being accelerated by a shock that is moving outwards. We estimated using DSA relations that this event is characterised by a Mach number $\mathcal{M}=1.95\pm0.01$, as inferred from the radio spectral index at the leading edge of the emission. No specific trend was obsereved in the southern relic. In this radio source we estimated a median $\alpha=-1.30$.
We then analysed the thermal and non-thermal spatial correlation by performing a point-to-point analysis of the X-ray and radio emission of the cluster. We found a sub-linear correlation between the $I_R$ and the $I_X$ of the halo region. This trend is consistent with current analysis on the X-ray and radio correlation in radio halos.  
 We investigated also the relationship between the thermal and non-thermal spatial correlation also in the northern radio relic. We found that an anti-correlation between $I_R$ and $I_X$ is present. A straightforward explanation of this behaviour could be related to the radial distribution of the X-ray and radio emissions. On one hand, the brightest radio emission of the relic is its leading edge, which is the most distant region from the cluster centre. On the other hand, the X-ray emission decreases with the distance from the centre and it is thus expected to be low (and to continue to decrease) in the relic regions. 
Finally, we compare the residuals obtained by dividing the X-ray and radio images to a double $\beta$-model and an exponential model to the X-ray and radio emission, respectively. We found no correlation between these two quantities. However, by comparing the X-ray residual and the radio emission of the halo, we found a strong correlation. This correlation is even higher than the one between X-ray and radio SB, suggesting that the process responsible for radio emission and the one that leaves fluctuations in the X-ray images are connected. Further investigations are needed to understand which are the physical processes that originate this behaviour, and to constrain the impact of the adopted X-ray emission model on this correlation.

\begin{acknowledgements}
We thank the staff of the GMRT who have made these observations
possible. The GMRT is run by the National Centre for Radio Astrophysics of the Tata Institute of Fundamental Research. S.E., L.L., F.G., M.R, acknowledge financial contribution from
the contracts ASI-INAF Athena 2019-27-HH.0, “Attività di Studio per la
comunità scientifica di Astrofisica delle Alte Energie e Fisica Astroparticellare” (Accordo Attuativo ASI-INAF n. 2017-14-H.0), INAF mainstream project
1.05.01.86.10, and from the European Union’s Horizon 2020 Programme under
the AHEAD2020 project (grant agreement n. 871158) AI acknowledgesfunding from the European
Research Council (ERC) under the European
Union's Horizon 2020 research and innovation
programme (grant agreement No. 833824). RJvW acknowledges support from the ERC
Starting Grant ClusterWeb 804208. MB acknowledges support from the Deutsche
Forschungsgemeinschaft under Germany's
Excellence Strategy - EXC 2121 “Quantum
Universe” - 390833306. GB acknowledges support from PRIN INAF
Mainstream Galaxy cluster science with LOFAR. A.I. acknowledges funding from the European
Research Council (ERC) under the European Union’s Horizon 2020 research
and innovation programme (grant agreement No. 833824').
\end{acknowledgements}

%
%

\bibliographystyle{aa}
\bibliography{biblio}

\appendix
\section{Error maps}\label{Appendix:errormap}
\begin{figure*}[h!]
    \centering
    \includegraphics[scale=0.7]{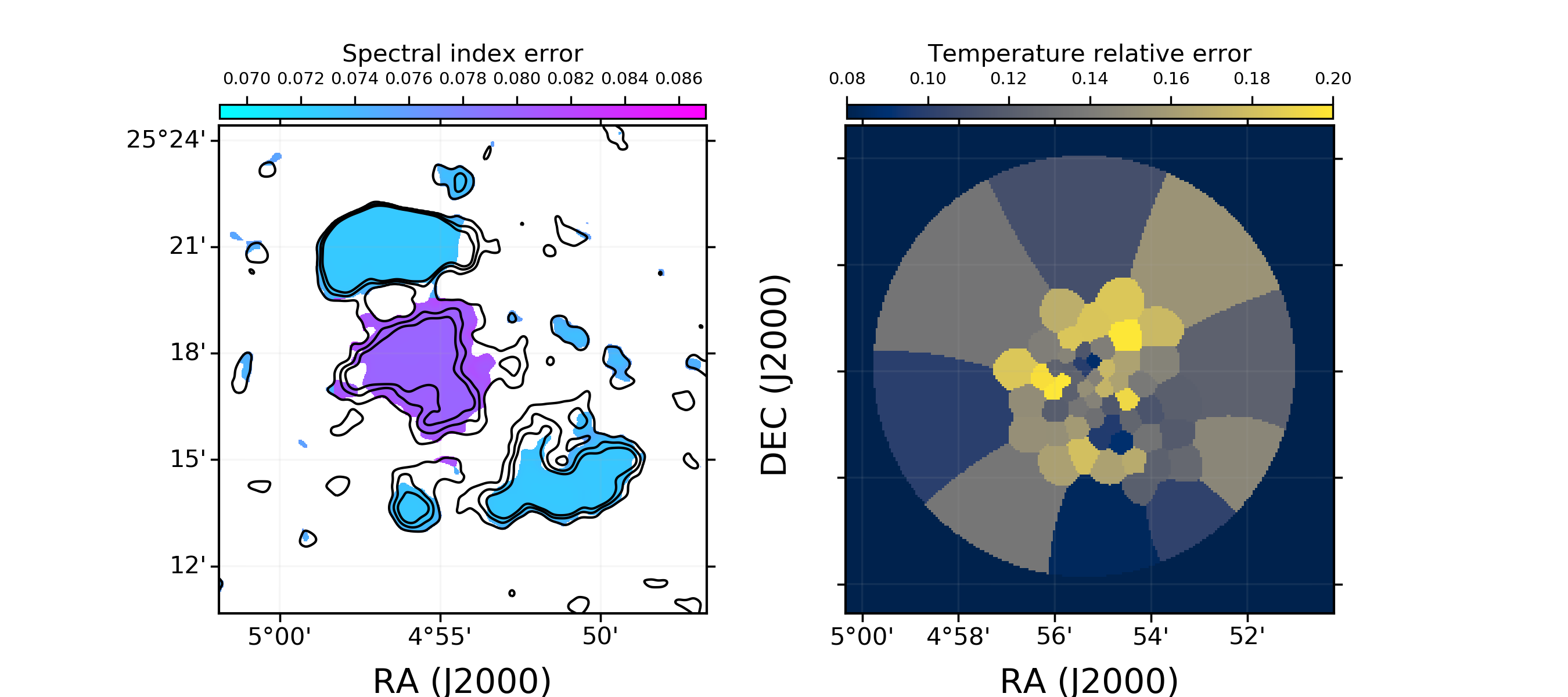}
    \caption{Left: spectral index error map. Right: temperature relative error map.}
    \label{fig:errmap}
\end{figure*}
We report in Fig.\ref{fig:errmap} the spectral index error and the temperature relative error map.

\section{High resolution map}\label{Appendix:pointsource}
\begin{figure*}
    \centering
    \includegraphics[scale=0.65]{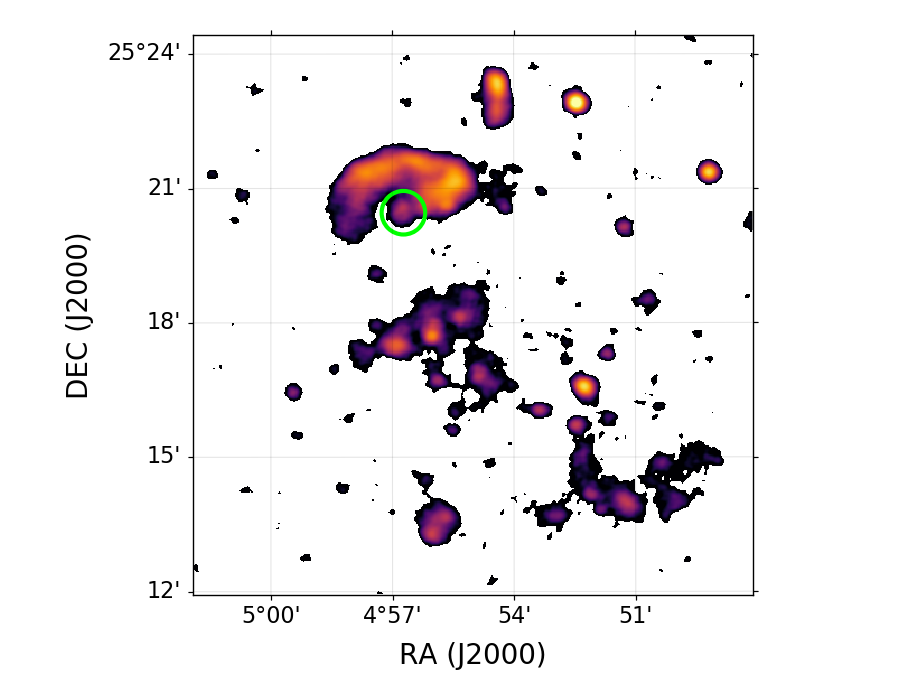}
    \caption{High resolution LOFAR image of the cluster G113.91. The beamsize and the rms of the map are respectively 15"$\times$15" and $\sigma=0.20$ Jy/beam. Pixels with values lower than 2$\sigma$ were masked. The green circle highlight the presence of the radio source that contaminates the estimate of $\alpha$ in the spectral profile of the norther relic.}
    \label{fig:pointosurcemap}
\end{figure*}

We report in Fig. \ref{fig:pointosurcemap} a hig resolution LOFAR map of the cluster G113.91. The circle shows the position of the radio source that produces an increase of $\alpha$ in the southern boundary of the northern relic.

\section{Correlation between the X-ray residuals and the radio emission}
\begin{figure*}
    \centering
    \includegraphics[scale=0.6]{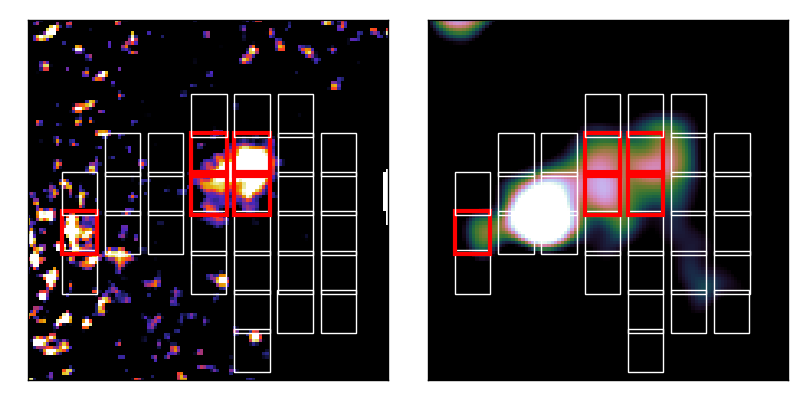}
    \caption{Left: residual of the X-ray emission obtained by dividing the observation for the model of the X-ray emission. The minimum of the colorscale is set to $\log(\text{data/model)=0.2}$. Right: LOFAR radio image of the halo region. The red cells of the grid corresponds to the five points with $\log (\text{data/model})>0.2$. } 
    \label{fig:redcells}
\end{figure*}

In Fig. \ref{fig:redcells} we show the location of the five points with  $\log\text{(data/model})>0.2$ that result from the plot of \ref{fig:correlazioni} (left panel). It is possible to observe that four of the five considered cells are located in correspondence of the X-ray core.

\end{document}